\documentclass[3p,times,twocolumn]{elsarticle}

\usepackage{graphicx}
\usepackage{amssymb}
\usepackage{amsmath}
\usepackage{slashed}
\usepackage{bm}
\newcommand{\be}{\begin{equation}}
\newcommand{\ee}{\end{equation}}
\newcommand{\bea}{\begin{eqnarray}}
\newcommand{\eea}{\end{eqnarray}}

\usepackage{amssymb}

\usepackage[figuresright]{rotating}

\begin{document}

\begin{frontmatter}




\title{The pseudo chiral magnetic effect in QED$_3$}


\author{Ana Julia Mizher$^1$, Alfredo Raya$^{2}$, Cristi\'an 
Villavicencio$^3$} 
\address{$^1$Instituto de Ciencias
  Nucleares, Universidad Nacional Aut\'onoma de M\'exico, Apartado
  Postal 70-543, M\'exico Distrito Federal 04510,
  Mexico.\\ 
  $^2$Instituto de F\'isica y Matem\'aticas, Universidad Michoacana de San 
Nicol\'as de Hidalgo, Ediﬁcio C-3, Ciudad Universitaria, C.P. 58040, Morelia, 
Michoac\'an, Mexico.\\
  $^3$Departamento de Ciencias B\'asicas, Universidad del B\'io-B\'io. Casilla 447,
Chill\'an, Chile.}

\begin{abstract}
Chiral magnetic effect (CME) has been suggested to take place during peripheral relativistic heavy ion collisions. However, signals of its realization are not yet independent of ambiguities and thus probing the non-trivial topological vacua of quantum chromodynamics (QCD) is still an open issue. Weyl materials, particularly graphene, on the other hand, are effectively described at low energies by the degrees of freedom of quantum electrodynamics in two spatial dimensions, QED3. This theory shares with QCD some interesting features, like confinement and chiral symmetry breaking and also possesses a non-trivial vacuum structure. In this regard, an analog of the CME is proposed to take place in graphene under the influence of an in-plane magnetic field in which the pseudo-spin or flavor label of charge carriers is participant of the effect, rather than the actual spin. In this contribution, we review the parallelisms and differences between the CME and the so-called pseudo chiral magnetic effect, PCME.     
\end{abstract}


\end{frontmatter}



\section{Introduction}

Nowadays, we are in an advanced stage about the knowledge of quantum chromodynamics (QCD) and the possibility to reach higher energies in order to explore in more detail heavier fundamental particles. 
However, one  intriguing aspect of the QCD theory is still missing in terms of experimental signals: its nontrivial vacuum structure. 
The possibility to detect some signal of topological effects due to the QCD vacuum sector seemed to be plausible when the chiral magnetic effect (CME) was proposed as a possible phenomenon in experiments of relativistic heavy ion collisions~\cite{CME1,CME2}. 
The CME is assumed to be produced in peripheral relativistic heavy ion collisions.  
Considering that at high temperature the plasma can present domains of metastable states where CP is violated by the topological configurations of the gauge sector. This symmetry breaking ends up manifesting itself as an imbalance of  quarks with different chiralities. The huge magnetic field generated during these processes due to the moving ions align the spins of the particles and produces an electric current \emph{along} the field lines. Until now, the possible signals of the CME \cite{data_CME} are controversial~\cite{signals}.
Nevertheless, there is an opportunity to study some features of high temperature QCD with novel materials in table top experiments. 
In Weyl materials,  particularly in graphene~\cite{graphene}, charge carriers are effectively described as Dirac particles in (2+1)-dimensions. 
The link between heavy ions in (3+1)-dimensions at high temperature and two dimensional cold materials is realized through quantum electrodynamics in (2+1)-dimensions (QED$_3$). 
In this work, we present a possible scenario  that mimics the CME in graphene, the so-called pseudo chiral magnetic effect (PCME)~\cite{colibri},  and explore the extent at which the main features of this effect can be translated as an analog behavior of QCD.

\section{Effective model of graphene}

Graphene~\cite{graphene} is a novel material with remarkable properties for technological applications, but which at the same time offers the possibility of exploring fundamental physics within table top experiments. It consist of a one atom thick membrane of carbon atoms tightly packed in an hexagonal array. The crystalline structure of single layer graphene has a bi-partite unit cell which permits to write the wavefunction as a two component object. At low energies, the nearest neighbors tight-binding description results in a linear dispersion relation near each of the two inequivalent Dirac points in the Brillouin zone, such that in the continuum, the underlying Lagrangian for the quasiparticle charge carriers corresponds to a free massless Dirac theory in two spatial dimensions with four component spinors, in which external electromagnetic fields enter through minimal coupling~\cite{gusynin:rev}, i.e., graphene is the incarnation of QED$_3$ in condensed matter physics.  Each bi-spinor in this theory describes the low-energy dynamics around the Dirac points, but rather than referring to the real spin of charge carriers, it is connected to a pseudo-spin or flavor for each triangular sublattice of the honeycomb array. 

The four-component spinors realization of QED$_3$ is interesting on its own. Two chiral-like transformations can be constructed from the matrices $\gamma^3$ and $\gamma^5$, because these do not enter into the dynamics. Thus, we are allowed to consider, besides the ordinary Dirac mass term, a Haldane mass term~\cite{haldane} $m_o\  \bar{\psi}[\gamma^3,\gamma^5]/2 \ \psi$, which breaks Parity and Time Reversal, and thus induces radiatively a Chern-Simons term for the gauge fields. Physical realization of Haldane mass term corresponds to distortions on the crystal structure of graphene, like strains~\cite{strain}. Consequently, the two sublattices are no longer equivalent and charge carriers are separated according to their pseudo-chiralities. To further connect
this observation with the CME, we consider an external magnetic field aligned along the graphene membrane described through the vector potential $A^{\mathrm{ext}}_3=Bx_2$, where $x_2$ repesents the second spatial coordinate along the graphene plane and we assume $B>0$. The above scenario can be described from  the Lagrangian
\begin{eqnarray}
{\cal L}_F &=& \bar\psi\big[i\slashed{D}
+(eA^{\mathrm{ext}}_3-m_3)\gamma^3  
-m_o\gamma^3\gamma^5\big]\psi\;,\label{lagr1}
\end{eqnarray}
where 
$D=(\partial_0-i\mu, v_F \bm{\nabla})$, $e$ is the fundamental charge, $v_F$ is the Fermi velocity, which from now onward we set to unity, and $\mu$ the chemical potential. The mass $m_3$ ensures the asymmetry between the sub-lattices. Experimentally, this can be realized by placing the graphene  membrane on top of a hexagonal boron nitride layer~\cite{BN}.
In the Weyl representation for the gamma matrices, the Lagrangian~(\ref{lagr1}) can 
be separated into two chiralities:
\begin{equation}
{\cal L}_F=  \sum_{\chi=\pm} 
\bar\psi_\chi\left[i\slashed{\partial}+\mu\gamma^0
+(eA^{\mathrm{ext}}_3 - m_\chi)\gamma^3
\right]\psi_\chi\;,\label{lagchiral}
\end{equation}
with  
$\psi_\pm=\frac{1}{2}(1\pm\gamma^5)\psi$ and $m_\pm = m_3\pm m_o$.
 
\section{Fermion Propagator}

Green functions can be obtained for each chirality separately, $G_\pm$, such that the total propagator can be defined as
\begin{equation}
G(x,x') = \frac{1+\gamma^5}{2}G_+(x,x')+\frac{1-\gamma^5}{2}G_-(x,x')\;.
\label{eq:prop}
\end{equation}
This Green function takes into account the presence of an external magnetic field, temperature and chemical potential. 
Finite temperature effects are introduced in the usual way by replacing the zeroth component of the fourth momentum as $k_0\to i\omega_n = i(2n+1)\pi T$ and momentum integrals $\int dk_0\to i2\pi T\sum_n$.
The inclusion of the external magnetic field can be treated with the Schwinger proper time method~\cite{schwinger}. 
However, when introducing the proper time in the presence of finite chemical potential some care must be taken, particularly with the range of integration in the proper time in order to guarantee the correct convergence of the integrals~\cite{chodos}. 
The generalization of this considerations can be simplified in terms of a regulator as 
\be
\int_0^\infty ds ~g(s) \to \int_{-\infty}^\infty ds\ r_s~ g(s), 
\ee
where the regulator reads
\begin{equation}
r_s(\omega_n\mu) = \theta(s)\theta(\omega_n\mu)\ {\rm e}^{-s\epsilon} - \theta(-s)\theta(-\omega_n\mu)\ {\rm e}^{s\epsilon}\;.
\end{equation}
The resulting propagator for each chirality is then
\begin{equation}
G_\pm(x,x') =    T\sum_n\int\frac{d^2 k}{(2\pi)^3}\ 
{\rm e}^{-ik\cdot(x-x')} ~\tilde G_n(k;\xi_\pm)\;,
\label{Gc3}
\end{equation}
with  
\begin{align}
 \tilde G_n(k;\xi_\pm)= i\int_{-\infty}^{\infty} \!\!\!
ds~r_s(\omega_n\mu)~
{\rm e}^{is{K_\parallel}^2-i
\left[{k_2}^2+\xi_\pm^2\right]
\frac{\tan(eBs)}{eB}}
\nonumber\\
\left\{ \slashed{K}_\parallel\left[1 \!+\! \gamma^2\gamma^3 \tan(eBs)\right]
+\left[k_2\gamma^2 \!+\! \xi_\pm\gamma^3\right]\sec^2(eBs)\right\},
\label{tildeG}
\end{align} 
$\xi_\pm =\frac{1}{2}(x^2+x'^2)eB+m_\pm$ and with  $K_\parallel = (i\omega_n+\mu,k^1,0)$. 
The Green function is nonlocal due to the term $\xi_\pm$.

\section{Observables for the PCME}

In order to establish the analogy between the PCME and the CME we calculate the electric (vector and axial) currents of each sublattice. Given the propagator in Eq.~(\ref{eq:prop}), the density of these currents are given by:
\bea \nonumber
j_\mu (x) &=& -e ~\langle \bar \psi \gamma_\mu \psi \rangle = e ~{\rm tr} \gamma_\mu G(x,x') \;, \\
j_{5\mu} (x) &=& -e~ \langle \bar \psi \gamma_\mu \gamma_5 \psi \rangle = e ~{\rm tr} \gamma_\mu \gamma_5 G(x,x').
\eea
The dependence on $y$, explicitly indicated, appears due to the nonlocal term in the chiral Green function. Tracing over gamma matrices, one can note that the only non-vanishing components of the currents are $j_1$ and $j_{15}$. This is in entire agreement with the CME mechanism, where an electric current in the direction of the magnetic field is generated as a product of the interplay of the topological vacua and the magnetic field. 

\begin{figure}
\centering
\includegraphics[scale=0.46]{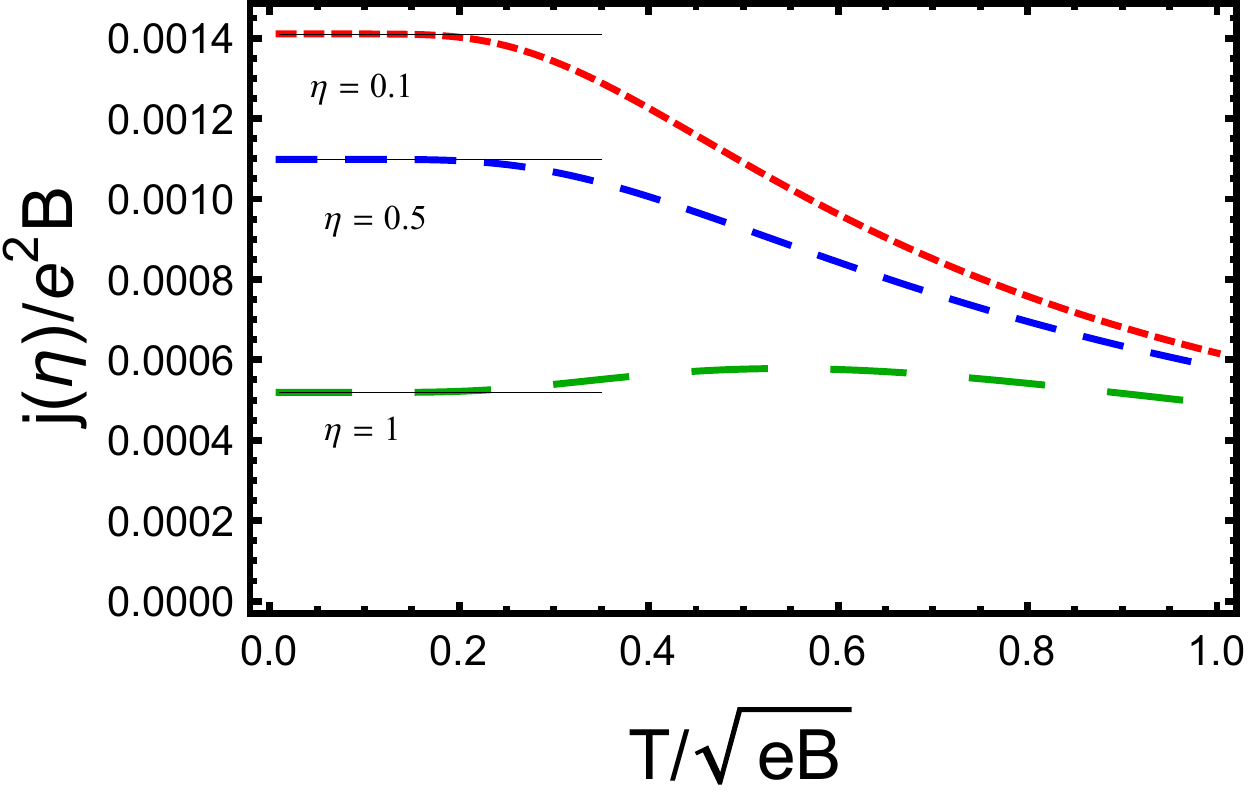}
\caption{Function $j$ (current density of each chirality) as a function of temperature (normalized to the appropriate unites of $eB$) for different values of $\eta$. The asymptotic behavior (dashed line) corresponds to the strong field limit, eq.~(\ref{eq:current_strongB}).}
\label{fig:current_density}
\end{figure}

To obtain the net contribution from both sublattices we trace over spin and pseudo-spin. Hence:
\bea \nonumber
j_1 \left(x^2\right) &=& j\left(x^2- x^2_+\right) - j\left(x^2-x^2_-\right),\\ 
j_{15} \left(x^2\right) &=& j\left(x^2- x^2_+\right) + j\left(x^2-x^2_-\right),
\eea 
with $x^2_{\pm}=-m_{\pm}/(eB)$ and the function $j$ is given by:
\bea \nonumber
j(\eta)=- \frac{e^2 B}{\pi}\sum_n \int_{-\infty} ^\infty ds\ r_s(\omega_n \mu)\  (\omega_n -i\mu) \\
 \left[\frac{\tan(eBs)}{eBs} \right] ^{1/2}\ {\rm e}^{-i\left(s(\omega_n -i\mu)^{2}+ eB\ \tan(eBs)\ \eta ^2 \right)}\;.
 \label{eq:current}
\eea
The proper time integral can be Wick-rotated as $s\to -is$ if $\mu<\pi T$. 
Then $\tan(eBs)\to -i\tanh(eBs)$.
Working in the strong field limit, one can consider $\tanh(eBs)\approx 1$, and the function $j$ simplifies to:
\be  
j(\eta)= \frac{e^2 B\ T}{\pi} \sum _n \int_{-\infty}^{\infty} \frac{ds}{(eBs)^{1/2}} {\rm e}^{-s(\omega_n -i\mu)^2 -eB ~\eta^2}.
\label{j_eta}
\ee
We can write conveniently  $s^{-1/2}= \pi^{-1/2}\int_{-\infty}^\infty dp e^{-sp^2}$, then, the integral in $s$ and the Matsubara frequencies can be performed easily.
After integration over $p$, Eq. (\ref{j_eta}) is reduced to the simple expression:
%
%
%
\be 
j(\eta)=4\frac{e\sqrt{eB}}{\pi^{3/2}} \mu {\rm e}^{-\eta ^2}.
\label{eq:current_strongB}
\ee 
It is interesting to notice that the final expression for the strong field limit is independent of the temperature. 
In Fig.~\ref{fig:current_density} we compare the expression obtained in Eq.(\ref{eq:current_strongB}) to the full numeric calculation of Eq.~(\ref{eq:current}) and show that they are in complete agreement for small values of the ratio $T/\sqrt{eB}$, which corresponds to large values of $eB$; the function $j$ assumes a $T$-independent value consistent with Eq.~(\ref{eq:current_strongB}).

\begin{figure}
\centering
\includegraphics[scale=.46]{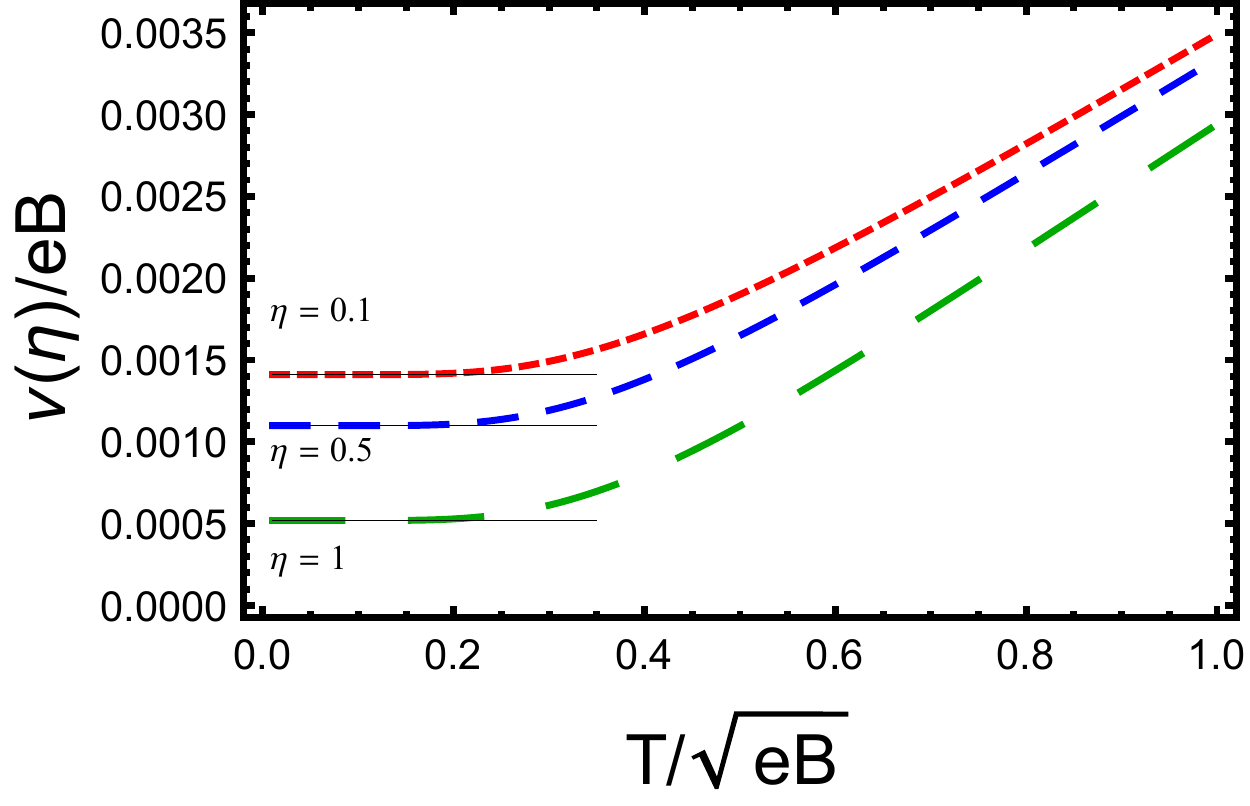}
\caption{Number density as a function of temperature (normalized to the appropriate units of $eB$) for different values of $\eta$. The asymptotic behavior (dashed line) corresponds to the strong field limit and matches the expression eq.~(\ref{eq:current_strongB}), as stated in the text. The scale of the plot is set by $eB=1$ and $e=1$.}
\label{fig:density}
\end{figure}

Performing a similar calculation for the number density and the chiral density number, defined as $n=\langle \psi^\dag \psi \rangle $ and $n_5=\langle \psi^\dag \gamma_5 \psi \rangle$, respectively. 
The densities can be, as in the case of the currents, written in terms of the number density of each chirality:
\begin{eqnarray}
n\left(x^2\right) &=& \nu\left(x^2-x^2_+\right)+\nu\left(x^2-x^2_-\right)\;,
\nonumber\\
n_5\left(x^2\right) &=& \nu\left(x^2-x^2_+\right)-\nu\left(x^2-x^2_-\right)\;,
\end{eqnarray}
with the function $\nu$ defined as
\bea \nonumber
\nu(\eta)=- \frac{eB}{\pi}\sum_n \int_{-\infty} ^\infty ds\ r_s(\omega_n \mu)\  (\omega_n -i\mu) \\
 \left[\frac{1}{eBs\ \tan(eBs)} \right] ^{1/2}\ {\rm e}^{-i\left(s(\omega_n -i\mu)^{2}+ eB\ \tan(eBs)\ \eta^2 \right)}.
 \label{eq:density}
\eea
In the strong field limit,  after rotation of the proper time $s\to -is$ and taking $\tan(eBs)\to -i\tanh(eBs)\approx -i$, 
we obtain the same result as for the current density (up to a unit charge factor $e$, of course). In Fig.~\ref{fig:density} we show that our approximation is in agreement with the full expression in Eq.~(\ref{eq:density}) in this regime. 

To obtain the full current $J_1$, we integrate the current density along a plane with finite length on the $x^2$-direction \cite{Miransky:2015ava}:
\be 
J_{1} = \int _{-L_2/2}^{L_2/2} dx^2 \left[j\left(x^2-x^2_+\right) -j\left(x^2-x^2_-\right)\right],
\ee 
and a similar expression holds for the total chiral number $N_5$, namely,
\be 
N_{5} = \int _{-L_2/2}^{L_2/2} dx^2 \left[\nu_5\left(x^2-x^2_+\right)- \nu_5\left(x^2-x^2_-\right)\right].
\ee 
It is straightforward to see that  the expressions for the current and chiral number even at the level of their corresponding densities coincide, namely, 
\be
J_1 = e N_5,
\ee
which is in accordance with the expected relation~\cite{CME2} for the CME in the strong magnetic field regime.
%
%





\section{Discussion and Conclusions}

A link between condensed matter physics and particle physics is established by the Quantum Electrodynamics in (2+1) dimensions. This theory presents many properties similar to Quantum Chromodynamics and at the same time provides a good description for the continuum limit of tight-binding constructions with massless Dirac fermions, which makes it applicable to a variety of planar systems in material science physics. This link opens for the possibility of application of some sophisticated tools, far developed for application on QCD, to these condensed matter systems. Consequently, it suggests that some phenomena well known in one of the two scenarios mentioned could have an analogue in the other. Inspired by this suggestion, we propose an arrangement for a graphene layer that reproduces all the ingredients needed to conform the chiral magnetic effect, a well known mechanism proposed for relativistic heavy ion collisions.  

Our construction is experimentally feasible. Essentially it involves a deformation of the layer in such a way that the equivalence between sublattices is broken - which is realizable by placing the graphene sheet over a substrate, as described in Ref.~\cite{BN} - and an in-plane magnetic field. Our results indicate a behavior very similar to the one in the QCD chiral magnetic effect: the generation of a conserved current aligned with the magnetic field as a product of the coupling between the field and the pseudo-spin associated to each sublattice in the graphene. Furthermore, in the limit of strong magnetic field we obtain that the current density is proportional to the chiral density number, which is in accordance with the QCD case. 

The effect we propose constitutes a new transport mechanism in graphene-like materials and provides a novel way to probe topological properties of the vacuum of gauge theories, connecting QCD to a more controled experimental environment.

\bigskip
{\small
The Huitzil collaboration acknowledges J.~A. Helayel-Neto for valuable discussions.
We also acknowledge hospitality of UMSNH (Mexico) and UBB (Chile), where parts of 
this work were carried out, and La Porfiriana for the inspiration. AJM acknowledges CONACyT-Mexico
under grant number 128534 and SECITI/CLAF under grant 023/2014. AR acknowledges support from CIC-UMSNH  and CONACyT-Mexico
under grants  No.~4.22 and 128534 respectively. CV acknowledges support from
FONDECYT under grant numbers 1150847, 1130056 and 1150471.
 }






\end{document}